\documentclass[a4paper]{article} 
\usepackage{graphicx}
\def\baselinestretch{1.3}
\def\miss {\hspace{-0.5cm}\slash~~}
\def\glue{\tilde{g}}
\catcode`@=11 
\def \gsim{\mathrel{\mathpalette\@versim>}}
\def \lsim{\mathrel{\mathpalette\@versim<}}
\def \@versim#1#2{\lower0.4ex\vbox{\baselineskip\z@skip\lineskip\z@skip
     \lineskiplimit\z@\ialign{$\m@th#1\hfil##\hfil$%
     \crcr#2\crcr\sim\crcr}}}
\catcode`@=12 

\begin{document}

\begin{flushright}
{ HRI-RECAPP-2009-002}
\end{flushright}

\begin{center}

{\large\bf Signals of supersymmetry with inaccessible first two families 
at the Large Hadron Collider}\\[15mm]
Nishita Desai\footnote{E-mail: nishita@mri.ernet.in}
and Biswarup Mukhopadhyaya\footnote{E-mail: biswarup@mri.ernet.in}\\
{\em Regional Centre for Accelerator-based Particle Physics \\
     Harish-Chandra Research Institute\\
Chhatnag Road, Jhunsi, Allahabad - 211 019, India}
\\[20mm] 
\end{center}

\begin{abstract} 
We investigate the signals of supersymmetry (SUSY) in a scenario where
only the third family squarks and sleptons can be produced at the
Large Hadron Collider (LHC), in addition to the gluino, charginos and
neutralinos. The final states in such cases are marked by a
multiplicity of top and/or bottom quarks.  We study in particular, the
case when the stop, sbottom and gluino masses are near the TeV scale
due to which, the final state t's and b's are very energetic.  We
point out the difficulty in b-tagging and identifying energetic tops
and suggest several event selection criteria which allow the signals
to remain significantly above the standard model background.  We show
that such scenarios with gluino mass up to 2~TeV can be successfully
probed at the LHC.  Information on $\tan \beta$ can also be obtained
by looking at associated Higgs production in the cascades of accompanying
neutralinos.  We also show that a combined analysis of event rates in
the different channels and the effective mass distribution allows one
to differentiate this scenario from the one where all three sfermion
families are accessible.

\end{abstract}

\vskip 1 true cm

\newpage
\setcounter{footnote}{0}

\def\baselinestretch{1.5}

\section{Introduction}

The investigation on whether nature is supersymmetric is an important
part of activities related to the Large Hadron Collider (LHC). By and
large, if supersymmetry
(SUSY)~\cite{Haber:1984rc,Martin:1997ns,Nilles:1983ge}, broken within
the TeV scale, has to offer a cold dark matter candidate, experiments
at the LHC should see signals with large missing transverse energy
($E_T \miss$), carried away by the lightest SUSY particle (LSP). The
lightest neutralino turns out to be the
LSP~\cite{Girardello:1981wz,Cremmer:1982vy} in most models.  Hard
leptons and/or jets of various multiplicity constitute the
accompanying `visible' signals when one has a neutralino LSP. It is
from these, then, that one is left to guess the detailed character of
the SUSY spectrum, and whether the low-energy spectrum is resulting
from some organising principle at high scale\cite{Nath:1983fp}.

A scenario often suggested is that the first two families of squarks
and sleptons are far too heavy ($\sim$ 5 - 10 TeV or more) to have any
impact on TeV-scale phenomenology, while the third family is within or
around a TeV in mass. While this still suffices in controlling the
quadratically divergent contributions to the Higgs mass, the
troublesome issue of flavour-changing neutral currents
\cite{Nir:1993mx} is avoided through decoupling of the first two
families~\cite{Brax:2000ip,Dimopoulos:1995zw}.  This kind of a SUSY
spectrum therefore deserves special attention in the context of the
LHC.  The present work suggests some improved criteria from which one 
not only obtains background-free signals of such a scenario, but also can
distinguish it from one where all three families of sfermions are
within the reach of the LHC.

Several theoretical schemes to achieve the suggested scenario have
been proposed in the literature. It is possible, for example, to have
a hidden sector of such composition that the third family couples to
it differentially, leading to smaller soft SUSY breaking terms
compared to those of the first two~\cite{Cohen:1996vb}. In particular,
such possibilities can be envisioned in string-inspired models with
flavour-dependent interactions with modular
fields~\cite{Barger:1999iv}. The existence of a horizontal symmetry,
with the third, and first two families being respectively singlets and
doublets under it, can also cause a mass
splitting~\cite{Pomarol:1995xc}. In SO(10) SUSY Grand Unified
Scenarios (GUT), too, suitable D-terms for the fields belonging to
$\bar{5}$ and $10$ of SU(5) may lead to a mass hierarchy of the
suggested type, with appropriate adjustment of
parameters~\cite{Baer:1999md,Baer:2001vw}.  A similar mass separation
can also arise out of the D-terms of some additional (anomalous) U(1)
gauge symmetry~\cite{Binetruy:1996uv,Dvali:1996rj}.  Finally,
appropriate regions in the parameter space of minimal supergravity
(mSUGRA), with a universal scalar soft breaking mass term well above a
TeV, can lead to lower values of only the third family sfermions due
to the role of Yukawa couplings in the process of running down to the
electroweak scale~\cite{Feng:1999zg,Chattopadhyay:2000qa}.

As we have already stated, our purpose is to take a close look at the
{\em LHC signals} of a scenario where only the third family sfermions
are within an accessible range. With this in view, we have chosen a
few benchmark points in the parameter space, where masses of the first
two families evolves down from a relatively high mass parameter at
high scale. In contrast, masses for the third family and the two Higgs
doublets originate in a relatively lower high-scale parameter, thus
creating a hierarchy of the type sought after.  The absolute as well
as relative values of the stop and sbottom masses are decided by other
parameters of the theory including $\tan \beta$, the ratio of the
vacuum expectation values (vev) of the two Higgs doublets, which is
turn controls the mixing between the left-and right-chiral states.

Many useful studies on the collider phenomenology of similar scenarios
have taken place earlier as well as in the very recent past.  These
include studies in both non-SUSY \cite{Baur:2007ck,Frederix:2007gi}
and SUSY scenarios
\cite{Barger:2006hm,Kadala:2008uy,Han:2008gy,Das:2007jn}. However,
with the LHC within close range, many aspects of the detection of new
signals are being realised with increasing degree of
sensitivity~\cite{Aad:2009wy,Ball:2007zza}. The present study is aimed
to supplement and extend the existing ones, keeping some such
realisations in mind, and to demonstrate the viability of some
additional final states and event selection criteria. To be specific,
some aspects, on which we have improved on earlier works, are as
follows:

\begin{itemize}
  \item The signals suggested in earlier works often depend on the
    identification of multiple b's in the final state. When the mass
    range of accessible superparticles are about a TeV or well above
    that, a large fraction of the b's arising from their cascades are
    quite energetic. The efficiency of b-tagging, on the other hand,
    is optimum for the transverse momentum ($p_T$) range of 50 - 100
    GeV\cite{Aad:2009wy}.  Although the performance of b-detection
    devices have scope for improvement beyond this, we felt that it is
    profitable to {\em suggest signals with only those b's whose $p_T$
      lie in the optimal range}.

  \item The signals often involve three or four top quarks in the
    final state.  Some of these tops can be considerably
    boosted. Since very energetic jets acquire invariant masses
    amounting to 15 - 20\% of their energy through spreading, it is
    not unlikely that these top quarks be faked by some energetic
    central jets in a machine like the LHC. Besides, as has been
    pointed out in recent studies
    \cite{Barger:2006hm,Kadala:2008uy,Han:2008gy,Han:2008xb,Baur:2007ck,Frederix:2007gi,Das:2007jn},
    top detection in this scenario has a rather low efficiency.
    Therefore, we wish to suggest signals where the likely presence of
    several tops can be exploited, but the tops by themselves need not
    be identified.

  \item With both of the above points in mind, we have suggested
    signatures of SUSY with mass spectra of the aforementioned type,
    by looking for various combinations of b's and leptons in the
    final state. Specific event selection criteria, especially those
    pertaining to the leptons, have been proposed to eliminate
    backgrounds and enhance the discovery reach. We have also gone
    beyond earlier studies by suggesting that final states with the
    lightest neutral Higgs, produced in association, can make the
    events stand out as a reflection of the nature of the neutralino
    spectrum.

  \item It is also of interest to find out if the proposed signals
    enable one to distinguish a SUSY spectrum where only the third
    family is accessible, from one where first two are also within the
    production threshold.  We suggest an effort in this direction by
    comparing the event rates in various signal channels and also
    looking at kinematic distributions such as the scalar sum of the
    $p_T$'s of all particles.
\end{itemize}

It may useful to specifically mention the points on which we have gone
beyond the earlier works cited in
\cite{Barger:2006hm,Kadala:2008uy,Han:2008gy,Das:2007jn}. In
\cite{Das:2007jn} and \cite{Kadala:2008uy} for example, b-tagging has
been highlighted as the main criterion (with an emphasis on $\ge 3$
b-jets in \cite{Das:2007jn}).  We have, on the other hand, taken the
position that b's may not be efficiently tagged when they are very
hard, and recommended that we depend on them only when their $p_T$
lies in the range $50 - 100~GeV$. We suggest the use of leptons, with
specific kinematic characteristics, to make good for `lost' b's. We
have also underscored the reasons why tops, being often very
energetic, be better not reconstructed.

We outline our parameter choice for the benchmark points in section 2,
where the justification of our approach is also given by showing the
kinematic properties of tops and b's corresponding to these
points. Studies on different signals as well as the strategies adopted
for suppressing backgrounds are reported in section 3. Section 4
contains a discussion on how one can hope to distinguish such a
scenario from one where all sfermion families are produced at the
LHC. We summarise our study and conclude in section 5.

\section{Choice of benchmark points: motivation for the chosen~signals}

The minimal supersymmetric extension of the standard model has more
than a hundred parameters. These parameters can be related by the
supersymmetry breaking scheme. Since our study is essentially
phenomenological, we economise on the parameters by considering an
mSUGRA-like scheme, with the difference that high-scale squark and
sfermion masses are not same for all generations.

We take
$(m_0^{(1,2)},m_0^{(3)},m_{\frac{1}{2}},sign(\mu),A^{(1,2)},A^{(3)},tan\beta)$
viz. scalar masses for the first two generations of sfermions, scalar
mass for the third generation of sfermions, unified gaugino mass, sign
of the Higgsino parameter $\mu$, the unified trilinear coupling for
first two generations, the trilinear coupling for the third generation
and $tan\beta$, where $\beta$ is the angle between the VEVs of the two
Higgs doublets to be the free parameters.  The first two families of
squarks and sleptons are degenerate and have rather high masses
$(\sim~5~TeV)$ whereas the third generation has masses in the range of
$1-1.5~ TeV$. As a consequence, the first two generations of sfermions
decouple and we have enhanced production of tops and bottoms in the
final states.

The benchmark points chosen by us in the above setting are based on
the following considerations:

\begin{itemize}
\item Being able to probe situations where the tops and bottoms coming
  out of SUSY cascades are energetic enough, so that their
  identification efficiency can be suspect. 

\item The stops and sbottoms being within the reach of the LHC, going
  to values as high as possible, while there are appreciable numbers
  of events with an integrated luminosity of $300~fb^{-1}$.

\item A scan over the gluino mass almost up to the search limit at the
  LHC, for medium as well as high values of the third family squark
  masses.

\item A fair sampling of values of $\tan\beta$, the chosen values
  being 5, 10 and 40.
\end{itemize}

\begin{table}[htb]
\begin{center}
\begin{tabular}{|c|c|c|c|c|c|c|c|}
  \hline
  
  \textbf{Point} & \textbf{$tan \beta$} &\textbf{$m_{\frac{1}{2}}$} &
  \textbf{${m_0}^{(3)}$} & \textbf{$m_{\glue}$} &
  \textbf{$m_{\tilde{t_1}(\tilde{t_2})}$} &
  \textbf{$m_{\tilde{b_1}(\tilde{b_2})}$}\\
  
  \hline
  \textbf{1A} &10 &800 &800 &1918 &1124 (1403)& 1376 (1502)  \\
  \hline
  \textbf{1B} &10 &600 &1000 &1496 &856 (1130) & 1100(1283)\\
  \hline
  \textbf{1C} &10 &400 &1200 &1063 &623 (916) &892 (1153)\\
  \hline
  \textbf{2A} &5 &600 &1000 &1496 &842 (1130) &1100(1290) \\
  \hline
  \textbf{2B} &5 &400 &1200 &1063 &603 (916) &890(1160) \\
  \hline
  \textbf{3A} &40 &600 &1000 &1493 &856 (1065) &1024(1157) \\
  \hline
  \textbf{3B} &40 &400 &1200 &1058 &619 (819) &783(982)\\
  \hline
\end{tabular}
\caption{Third generation squark and gluino masses in GeV for the
  benchmark points considered.}

\vskip 0.25in

\begin{tabular}{|c|c|c|c|c|c|c|}
\hline 
\textbf{Point} & \textbf{$m_{\chi^+_1}$} &\textbf{$m_{\chi^+_2}$}
&\textbf{$m_{\chi^0_1}$} &\textbf{$m_{\chi^0_2}$}
&\textbf{$m_{\chi^0_3}$} & \textbf{$m_{\chi^0_4}$}\\
\hline
1A &660 &881 &348 &660 &864 &881 \\
\hline
1B &484 &648 &258 &484 &622 &649 \\
\hline
1C &288 &409 &167 &290 &356 &410 \\
\hline
2A &487 &707 &258 &488 &686 &701 \\
\hline
2B &313 &492 &168 &315 &465 &492 \\
\hline
3A &482 &619 &259 &482 &590 &619 \\
\hline
3B &261 &384 &166 &265 &302 &383 \\
\hline

\end{tabular}
\caption{Chargino and Neutralino masses in GeV for all the benchmark
  points.}
\end{center}
\end{table}

With this in mind, high scale value of ${m_0}^{(1,2)}$ is set to
$5~\mathrm{TeV}$ for the first two families, while the high-scale mass
for the third family ($m_0^(3)$) is set so as to obtain third
generation squark masses of the order of $1~TeV$.  The trilinear
couplings $A_i$ are all set to zero and we choose $\mu > 0$.  We
mostly focus an $tan \beta = 10$ but also look at $tan \beta = 5,40$
to see if any major differences are indicated.  The Higgs mass
parameters $M_{H_u}$ and $M_{H_d}$ are set to the value of the third
family $m_0$ at the high scale. 

The particle spectrum has been generated using SuSpect~2.34 using high
scale inputs in the pMSSM (phenomenological MSSM) option. The squark
and gluino masses for the various benchmark points are given in
Table~1. The masses for charginos and neutralinos are given in
Table~2.  The points itemised above, together with a glance at Tables
1 and 2, should convince the reader that the choice of our benchmark
points are broadly representative of the scenario investigated
here. It is obvious that in all these cases tops and bottoms will
populate the final state, but will be often carry very high energies.

We explore regions of the parameter space where squarks are lighter
than the gluino and of the order of $\sim 1~\mathrm{TeV}$.  Cases
where the gluino is considerably lighter than all squarks are left out
for the following reasons. First, such a situation is typical of a
focus point scenario, which has been already investigated
\cite{Das:2007jn}.  Secondly, the gluino in such cases will have
three-body decays only, and the tops and bottoms produced in the
process will not be excessively hard, so that the conventional search
strategies should work well. Thirdly, with a relatively light and
therefore copiously produced gluino, there can be like-sign dilepton
events in abundance, thus making the scenario conspicuous.

A b-tagging efficiency of $50\%$ with a rejection of QCD jets at more
than $99\%$ is well established for b-hadrons with the transverse
momentum $(p_T)$ between 50 to 80 GeV. But in our case, it can be seen
that the $p_T$ of b-hadrons very often exceeds this. It is not clear
how the efficiency goes down as $p_T$ increases above 100 GeV.  The
$p_T$-distribution of b's in four-b events can be seen in
Figures~\hbox{1 and 2}. \footnote{ While there are many events in our chosen regions
with both three-and four top quarks in the final state, 3b final states 
are only possible via squark-gluino production, and that too 
driven by the b-quark distribution in the proton. Thus the number of
3b events is relatively small.}

Top quarks can be identified by a combination of a b-jet and a $W$
which give an invariant mass within a window of the top mass . The
candidate $W$s are obtained from jet-pairs having invariant mass in
the range $M_W \pm 15~\mathrm{GeV}$.  Besides the aforementioned
b-tagging difficulty, this top reconstruction is complicated by two
other factors in our situation.

First, at very high boosts, the jets from decay of the top can be
highly collimated. However, very high energy QCD jets can also develop
an invariant mass up to 15 - 20\% of the jet energy, and thus, a top
depositing a large energy in the hadron calorimeter can be faked by a
similarly energetic jet whose `effective' invariant mass may be of the
same order as the top mass.  In such cases, one has to resort to
special techniques, such as specific kinematics, energetic leptons
contained in jets, and using jet-substructure. Such techniques have
been studied recently by various
groups~\cite{Kaplan:2008ie,Almeida:2008tp}.

Secondly, we have Higgs production through the cascade $\chi_2^0
\rightarrow h \chi_1^0$. The $\chi_2^0$ is produced in about 50\% of
the events we generate, and low $\tan \beta$ its decay into a Higgs
has a large branching ratio.  The Higgs then decays into a pair of
b's.  The mass of the Higgs in all our benchmark points lies at $\lsim
120~\mathrm{GeV}$. In cases where both the b-jets are not identified,
the W-peak from invariant mass of jet-pairs, which is important in
retracing the top via the W, is largely washed out by that of the
Higgs and due to the large combinatorial background arising from a
large jet multiplicity. Thus our benchmark points highlight one
additional difficulty in identifying the final states via the top.

To ameliorate these difficulties, we do not emphasise the
reconstruction of the top. We also supplement b-tagging by identifying
hard leptons from the decay of energetic top quarks. We find that
looking for leptons of various multiplicity can compensate for the
potentially low tagging efficiencies for very high energy b's.

We are looking at very high masses for squarks and gluinos and
consequently rather low production cross sections.  Thus, it will
require a large integrated luminosity at the LHC to achieve the
required statistical significance.  By that time, we assume that the
lightest neutral SUSY Higgs has already been identified. An additional
handle for our benchmark points is thus provided by the possibility of
looking for final states with leptons/b-quarks, together with not only
large missing energy but also a Higgs in the final state, identified
by a mass peak.

Thus our chosen benchmark points elicit a number of features of the
signals of SUSY with only the third family of sfermions accessible. We
use these in our study of the suggested signals in the next section.

\begin{figure}[tb]
\begin{center}
\includegraphics[width=70mm]{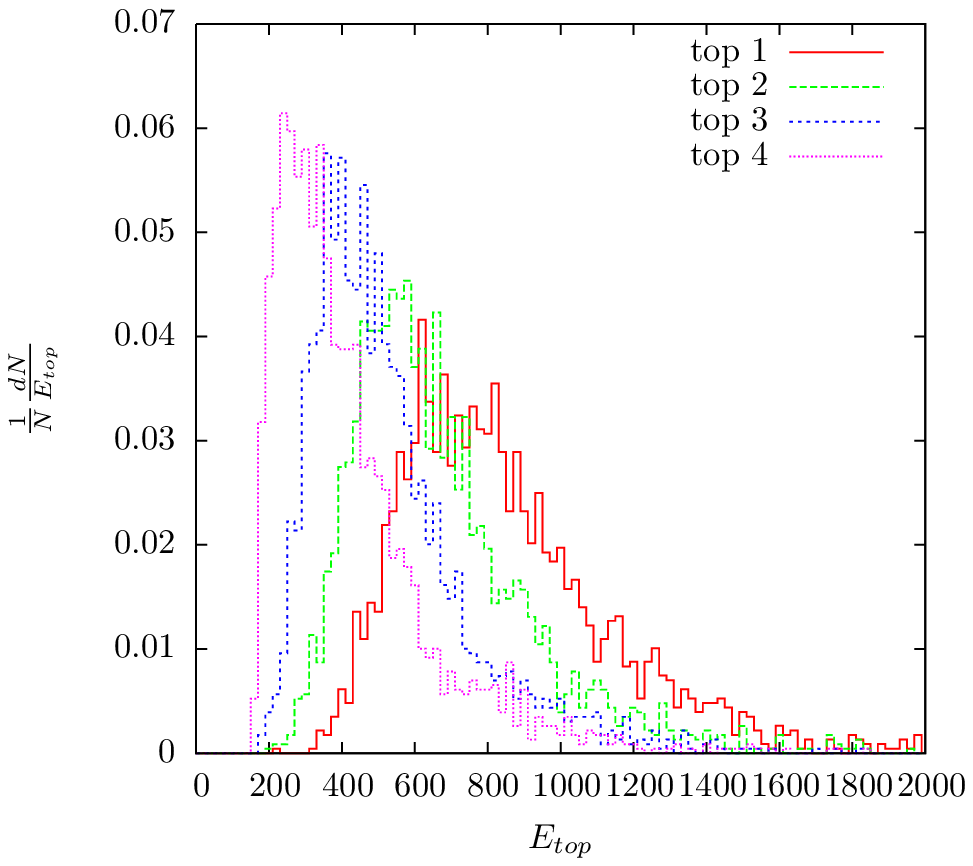}
\includegraphics[width=70mm]{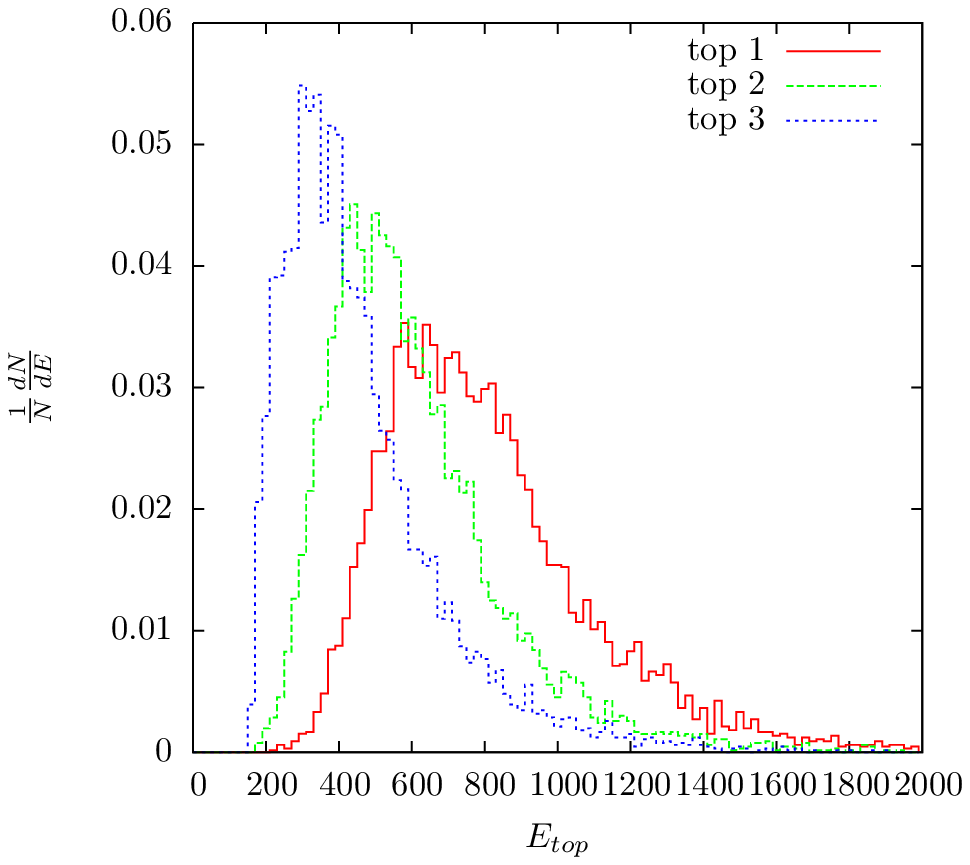}
\caption{Energy$(E)$ distribution of tops for four- and three-t events for benchmark point 1A.}

\vskip 0.20in
\includegraphics[width=70mm]{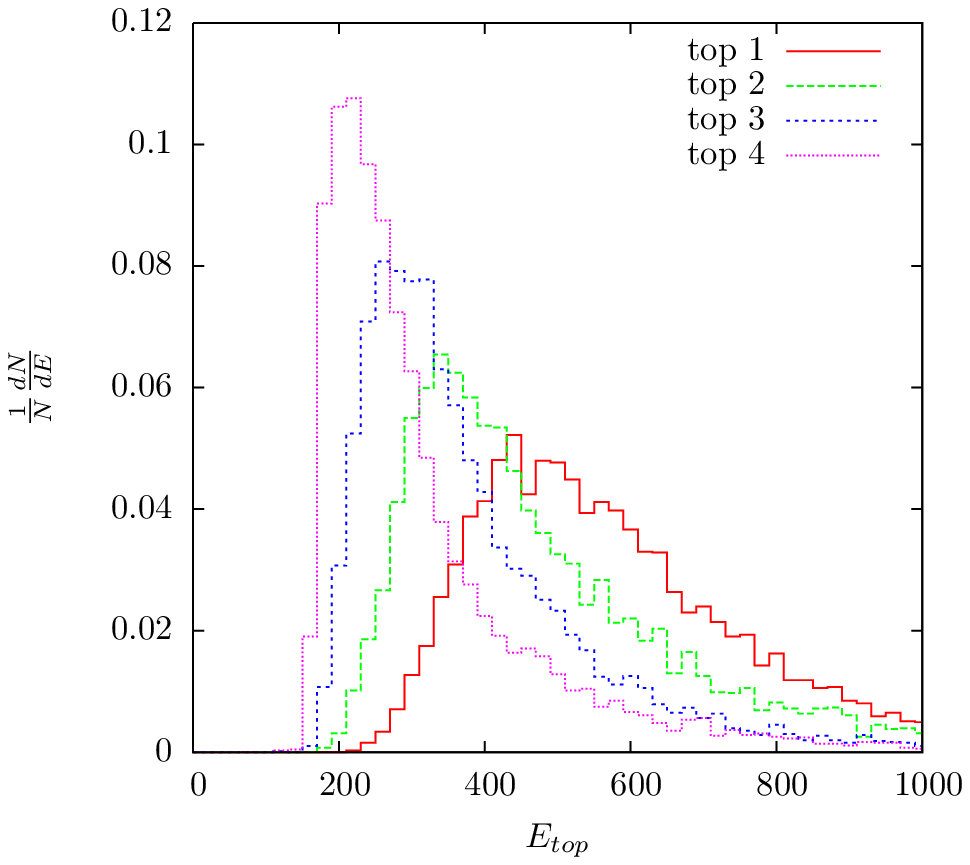}
\includegraphics[width=70mm]{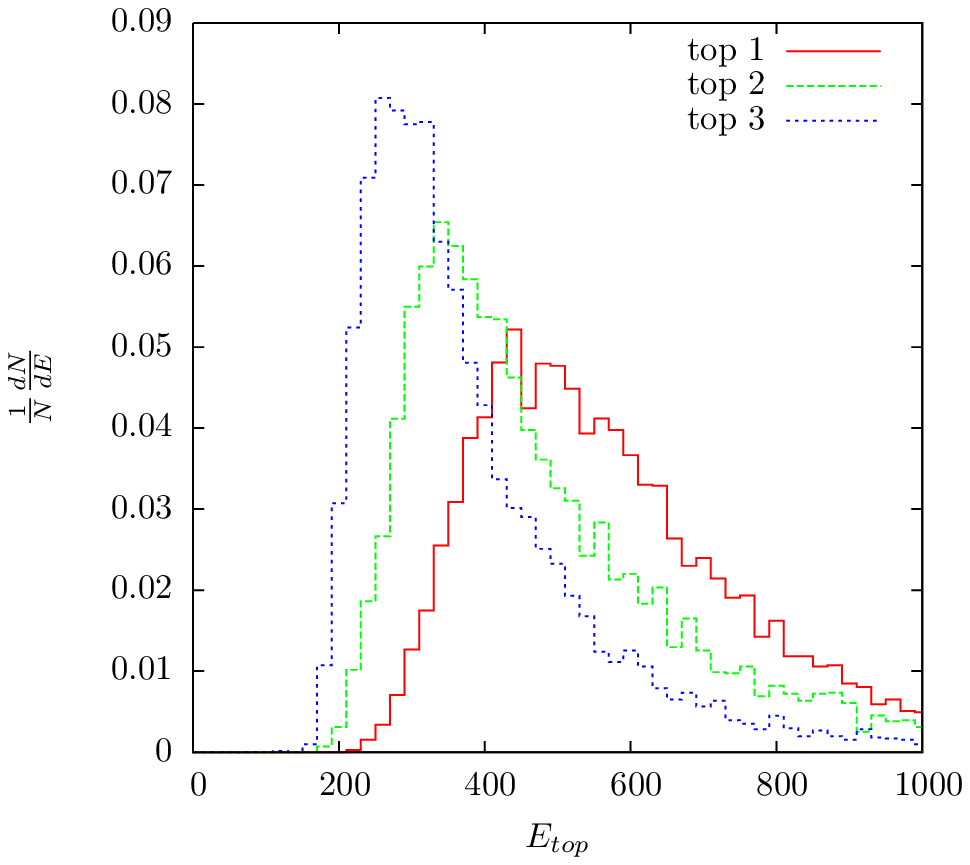}

\caption{Same as in Figure~1, for benchmark point 1C.}
\end{center}
\end{figure}

\begin{figure}[htbp]
\begin{center}
\includegraphics[width=70mm]{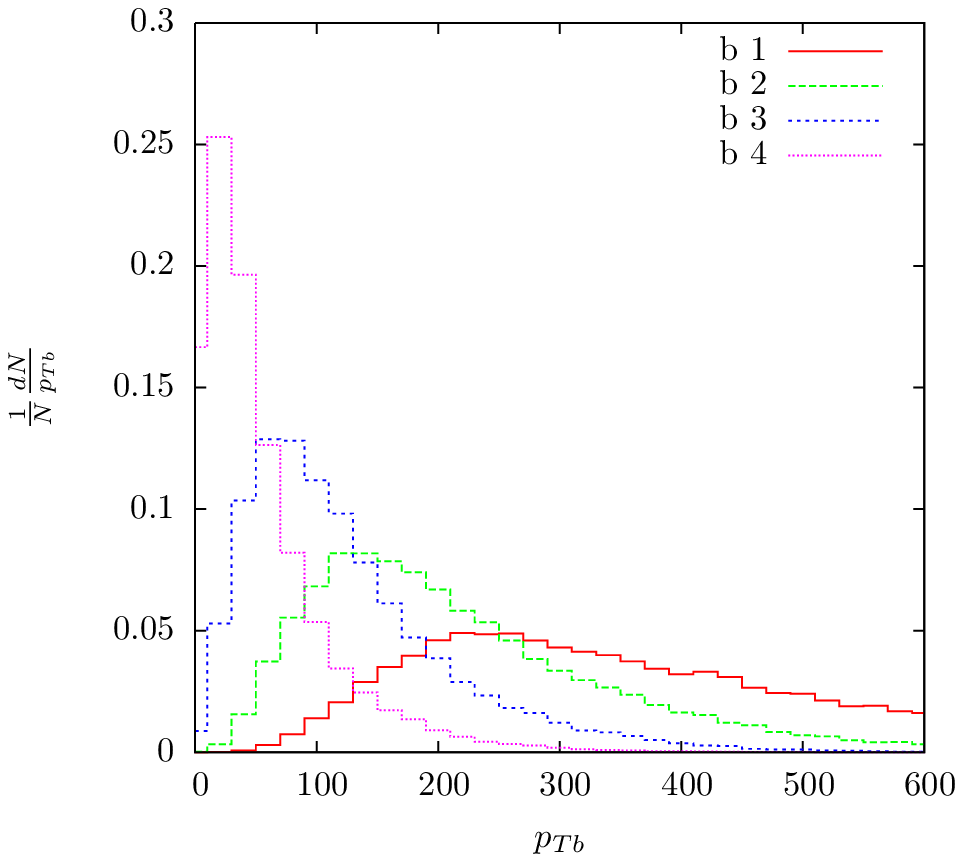}
\includegraphics[width=70mm]{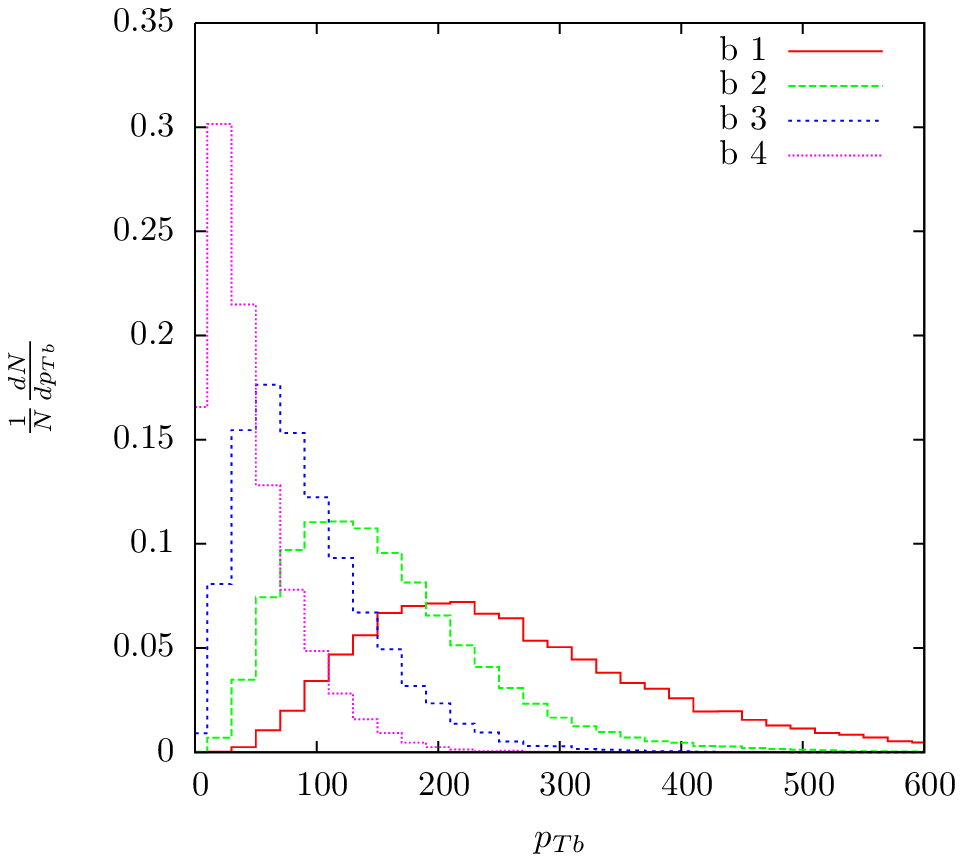}
\caption{Transverse momentum $p_T$ of b-quarks for points 1A and 1C.}
\end{center}
\end{figure}

\section{Signals and Backgrounds}

We are concerned primarily with observing final states with a large
number of top and bottom quarks.  Signal events have been generated
using Pythia v6.409 \cite{Sjostrand:2006za} by allowing the
squark-squark, gluino-gluino and squark-gluino production channels.
We have used CTEQ5L parton distribution functions. The factorisation
and renormalisation scales have been set to
$\mu_R=\mu_F=\sqrt{p_{\perp}^2+(P_1^2+P_2^2+m_3^2+m_4^2)/2}$ where
$P_1,P_2$ are the virtualities of the incoming particles, $p_\perp$ is
the transverse momentum of the scattering process and $m_3,m_4$ are
the masses of the outgoing particles in the initial hard scattering
process.

We concentrate on three and four-top events in particular.  The
$\tilde{g}$ can decay into $t\tilde{t}_{1,2}$ or
$b\tilde{b}_{1,2}$. Whenever it is kinematically allowed, the squarks
can then decay via $\tilde{t}_{1,2}\rightarrow t\tilde{\chi}_{i}^0$
(with $i=1-4$), $\tilde{t}_{1,2}\rightarrow b\tilde{\chi}_{1,2}^+$,
$\tilde{b}_{1,2}\rightarrow b\tilde{\chi}_{i}^0$ and,
$\tilde{b}_{1,2}\rightarrow t\tilde{\chi}_{1,2}^+$. Thus,
$\tilde{g}\tilde{g}$ production can give four-top final states via
$\tilde{g}\tilde{g}\rightarrow t\tilde{t}_{1,2} t\tilde{t}_{1,2}$ and
each $\tilde{t}_{1,2}\rightarrow t\tilde{\chi}_{i}^0$. Three-top final
states can be obtained when $\tilde{g}\tilde{g}\rightarrow
t\tilde{t}_{1,2} b\tilde{b}_{1,2}$ with $\tilde{b}_{1,2}\rightarrow
t\tilde{\chi}_{1,2}^+$.  Figures~1 and~2 give the energy distribution
of the top quarks for benchmark points 1A (highest squark/gluino
masses) and 1C (lowest squark/gluino masses). The transverse momentum
($p_T$) distribution of the b-quarks is shown in Figure~3.

As has been mentioned in the previous section, we have examined final
states with various combinations of b's and leptons. We comment first
on certain generic features of signal identification, before the
numerical results for each signal are presented.  These features also
help us in evolving the event selection criteria for this scenario.

\begin{itemize}

  \item \textbf{Identification of leptons $(e,\mu)$ :} We are
    interested in identifying leptons coming from top decay.  Since
    the parent W of the lepton is on-shell, we expect that the lepton
    to be well isolated from the nearest jet. We first identify
    leptons with the following cuts:
    \begin{enumerate}
      \item $p_T^l > 10~\mathrm{GeV}$ (trigger)
      \item Separation from each jet $\Delta R_{lj} >0.4$
    \end{enumerate}

    \noindent Lepton momenta are smeared according to the prescription
    $\sigma(E)=a\sqrt{E}+bE$ where $\sigma(E)$ is the resolution, with
    $a=0.055(0.02)$ and $b=0.005(0.037)$ for electrons (muons) and
    energy measured in GeV.

    \noindent We subsequently apply further cuts for each channel to
    restrict to leptons coming from tops.

  \item \textbf{Jets:} Jets are formed using the routine
    \texttt{PYCELL} built into \textsc{Pythia} The jet energy is
    smeared using $\sigma(E)=\sqrt{E}$. The parton-level processes
    that lead to the final states of interest to us have usually a
    large jet multiplicity. Using \textsc{Pythia}, the multiplicity
    peaks at 6 when both initial and final state radiation are taken
    into account.  With this in view, we have always demanded a
    minimum of four jets in the final state.

  \item \textbf{b-Tagging:} In the absence of any clear guideline on
    the tagging efficiency for very high-$p_T$ b-hadrons, we take a
    conservative approach and restrict our b-tagging capabilities to
    hadrons with $p_T$ between 50 and 100 GeV. A jet is assumed
    b-tagged with an efficiency of 0.50 if:
    \begin{enumerate}
      \item A b-hadron lies within a cone of $\Delta R < 0.5$ of the jet-axis
      \item The b-hadron has a $50~GeV \leq p_T \leq 100~\mathrm{GeV}$.
    \end{enumerate}

    \item \textbf{Missing transverse energy ($E_T\miss$) and the
      effective mass ($M_{eff}$):} Since we are considering R-parity
      conserving supersymmetry, the lightest supersymmetric particle
      (LSP) is stable.  In our case, the first neutralino is the LSP
      and since it is uncharged, it escapes detection. This gives a
      very large missing-$E_T$ which gives us the first handle for
      discriminating supersymmetric events.  Also, since the masses of
      the supersymmetric particles are very high for the scenarios
      investigated here, the effective mass of the event, defined by
      $M_{eff}=\sum_{jets} |p_T^j| + \sum_{leptons} |p_T^l| +
      E_T\miss$ also takes a very high value compared to what is
      expected of standard model processes.  The $E_T \miss$ and
      $M_{eff}$ distributions for two benchmark points are shown in
      Figure~4, along with the corresponding distribution for standard
      model backgrounds.

      The calculation of $E_T\miss$ has to take into account not only
      the `visible' $\mathbf{p}_T$ due to jets, leptons and photons that
      satisfy the requisite triggers but also objects with
      $p_T>0.5~\mathrm{GeV}$ and $|\eta|<5$ which are not identified
      as leptons or do not fall within any jet cone. The contribution
      from this extra part is summed up as the `soft-$p_T$' component.
      This is smeared according to the prescription
      $\sigma(p_T)=\alpha \sqrt{p_T}$ with $\alpha=0.55$.  The total
      visible transverse momentum is given by
      $\mathbf{p}_T^{vis}=\sum_{jets} \mathbf{p}_T^j + \sum_{leptons}
      \mathbf{p}_T^l + \mathbf{p}_T^{soft}$.  Missing $E_T$ is then
      the magnitude $|\mathbf{p}_T^{vis}|$.

\end{itemize}

\begin{figure}[htbp]
\begin{center}
\includegraphics[width=70mm]{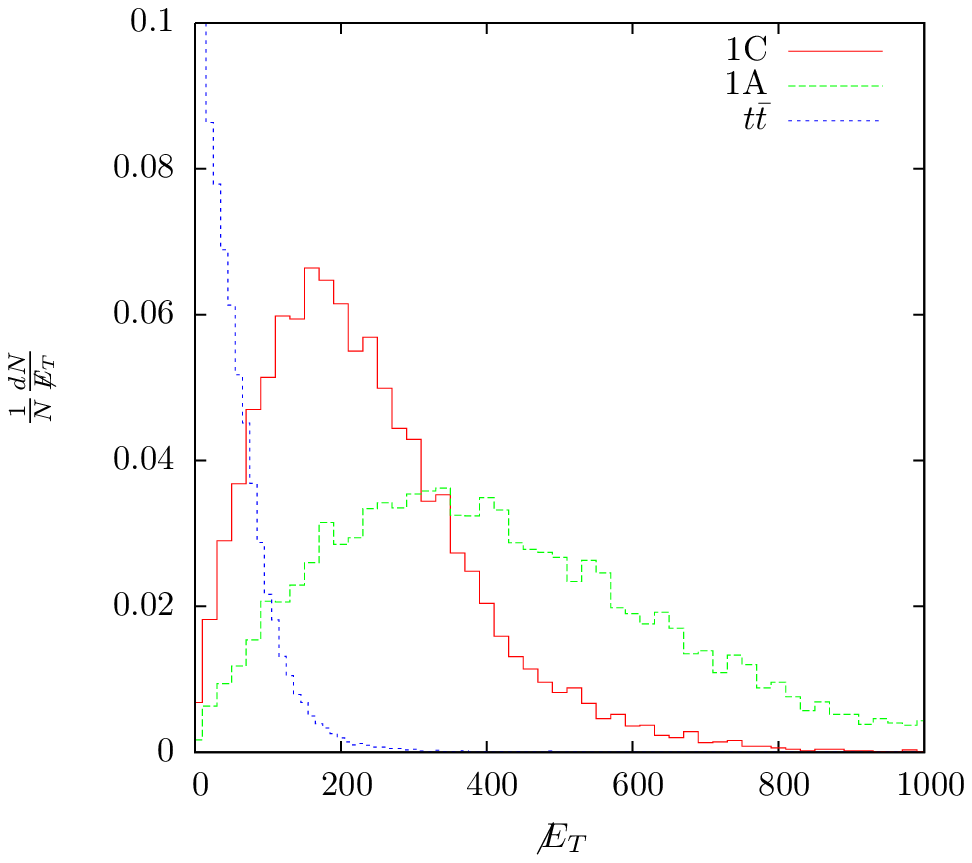}
\includegraphics[width=70mm]{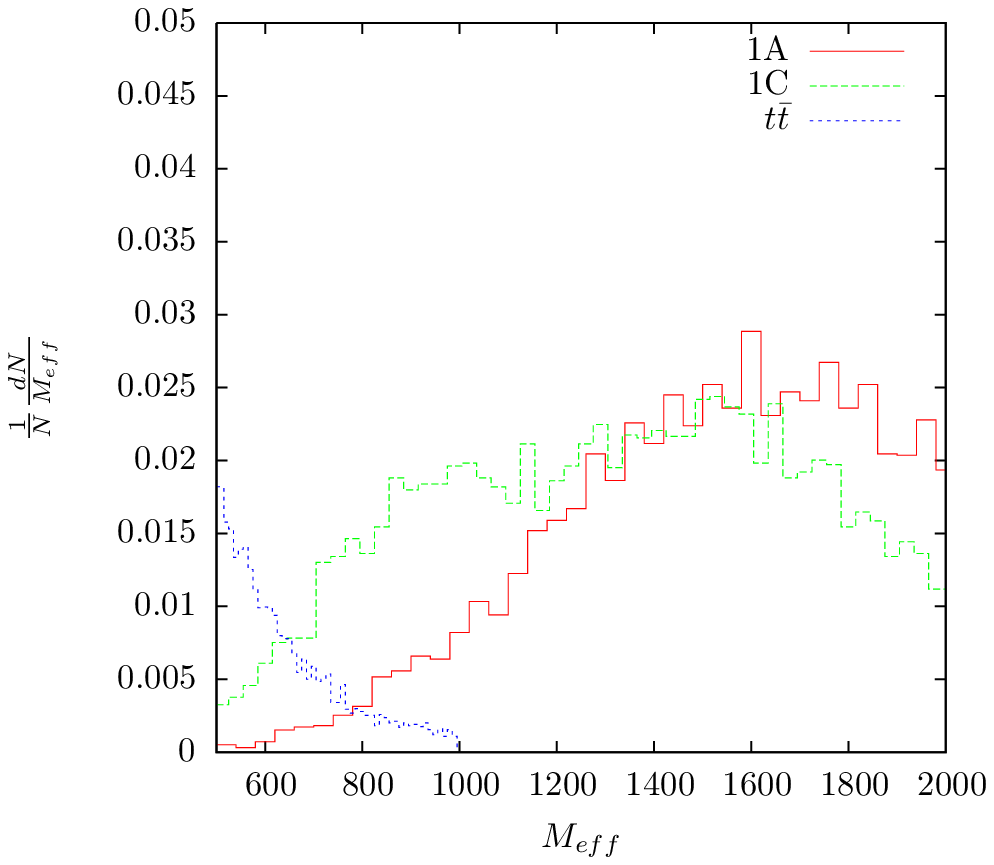}
\caption{Missing transverse energy ($E_T\miss$) and effective mass
  $M_{eff}$ distribution for benchmark points 1A, 1C and the dominant
  standard model background ($t\bar{t}$).}
\end{center}
\end{figure}

In gluino decay, the production of the $\tilde{\chi}_2^0$ occurs in
about $50\%$ of all events. For the benchmark points with $\tan
\beta=5,10$, the difference between masses of the second and the first
neutralino is more than the mass of the lightest neutral Higgs
($m_{h^0}$). The most common decay channel $\tilde{\chi}_2^0
\rightarrow h \tilde{\chi}_1^0$ yields a neutral Higgs in the final
state which then decays into a pair of b-quarks. This is because there
are two two-body decays of the $\tilde{\chi}_2^0$, namely,
$\tilde{\chi}_2^0 \rightarrow h \tilde{\chi}_1^0$, $\tilde{\chi}_2^0
\rightarrow Z \tilde{\chi}_1^0$ and the three body decay
$\tilde{\chi}_2^0 \rightarrow \tau\tilde{\tau}\tilde{\chi}_1^0$. Of
these, the third one is kinematically disallowed due to large
$m_{\tilde{\tau}}$. The decay into a Z is suppressed by the product of
Higgsino components of both $\tilde{\chi}_2^0$ and
$\tilde{\chi}_1^0$. The decay into a Higgs requires the Higgsino
component of any one neutralino and it therefore wins when kinematics
are favourable.  For the case with $\tan \beta=40$, the mass
difference $m_{\tilde{\chi}_2^0} - m_{\tilde{\chi}_1^0}$ is smaller
than $m_{h^0}$.  As a result, $\tilde{\chi}_2^0 \rightarrow Z
\tilde{\chi}_1^0$ is the dominant decay.  The identification of Higgs
can therefore give us information on the value of $\tan \beta$.
 
Based on the above observations, we now list the basic 
cuts that have to be satisfied by all events:
\begin{enumerate}
  \item $E_T\miss \geq 300~\mathrm{GeV}$
  \item$ m_{eff}=\left(\sum{|\vec{p_T}|}+E_T\miss\right) \geq 1000~GeV$
  \item Jet multiplicity $n_{jet} \geq 4$
  \item $p_T(j_1)>100~\mathrm{GeV}$
  \item $p_T(j_2)>80~\mathrm{GeV}$
  \item $p_T(j_3)>40~\mathrm{GeV}$
\end{enumerate}

\noindent
The inclusive cross sections for `all events' satisfying the basic
cuts for our benchmark points are summarised in Table~3.

\begin{table}[tbh]
\begin{center}
\begin{tabular}{|c|c|c|c|c|c|c|c|}
  \hline
  \textbf{Point} & \textbf{1A} & \textbf{1B}& \textbf{1C} &\textbf{2A} &\textbf{2B} &\textbf{3A} & \textbf{3B} \\
  \hline
  \textbf{$\sigma_{no cuts}$} &4.51 &32.47 & 308.00 & 37.07 &352.01 &34.62.0 &337.51    \\
  \hline
  \textbf{$\sigma_{basic}$}   &3.89 &15.09 & 83.87  & 17.21 &98.31  & 16.62  &93.767  \\
  \hline
\end{tabular}
\caption{Total $\tilde{g}\tilde{g}$, $\tilde{g}\tilde{q}$ and
$\tilde{q}\tilde{q}$ production cross sections for all the benchmark
points before and after basic cuts.}
\end{center}
\end{table}

We now discuss signals in various channels.  The cuts or extra
identification criteria applied henceforward will be over and above
the basic cuts enumerated above.

\subsection{Channels: $1b+2l$, $1b+2l_{(SSD)}$ and $2l_{(SSD)}$}

As mentioned earlier, the tops produced from the decay of heavy
squarks and gluinos are highly energetic.  Even in three-top
(four-top) events which would give three (four) b-quarks, it is not
always possible to tag all of them.  However, we expect that leptons
arising out of the decay of the tops to be very energetic.  Therefore,
we look at two energetic leptons with and without additional b-tags.

The backgrounds are calculated including the processes
$t\bar{t}+jets,Wb\bar{b}+jets,Wt\bar{t}+jets,Zt\bar{t}+jets,Zb\bar{b}+jets,
4t,4b$ and $2t2b$ generated with the help of
ALPGEN~\cite{Mangano:2002ea}.  Most of the background comes from the
$t\bar{t}$ channel. The $p_T$ distributions for leptons for benchmark
points $1A$ and $1C$ along with $t\bar{t}$ are given in Figure~6.  We
therefore apply the following cuts to select leptons over those from
standard model backgrounds.

\noindent The final cuts on the leptons are:
\begin{enumerate}
  \item $p_T(l_1) \geq 80~\mathrm{GeV}$
  \item $p_T(l_2) \geq 30~\mathrm{GeV}$
\end{enumerate}

\begin{figure}[h]
\begin{center}
\includegraphics[width=70mm]{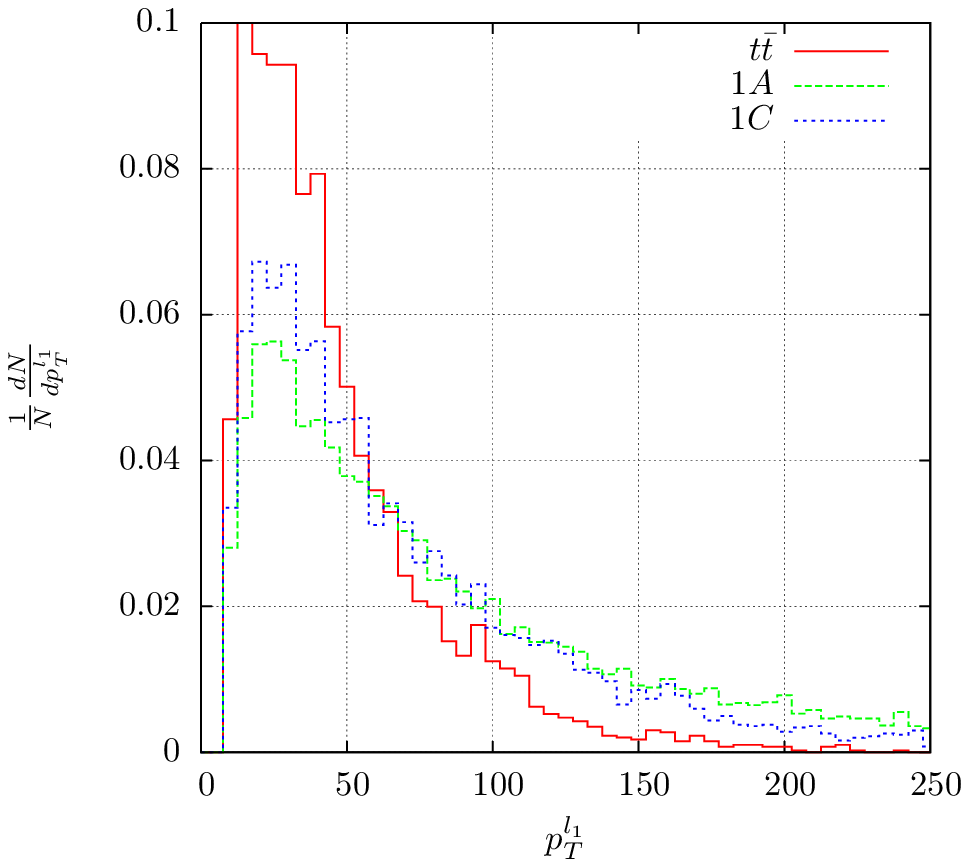}
\includegraphics[width=70mm]{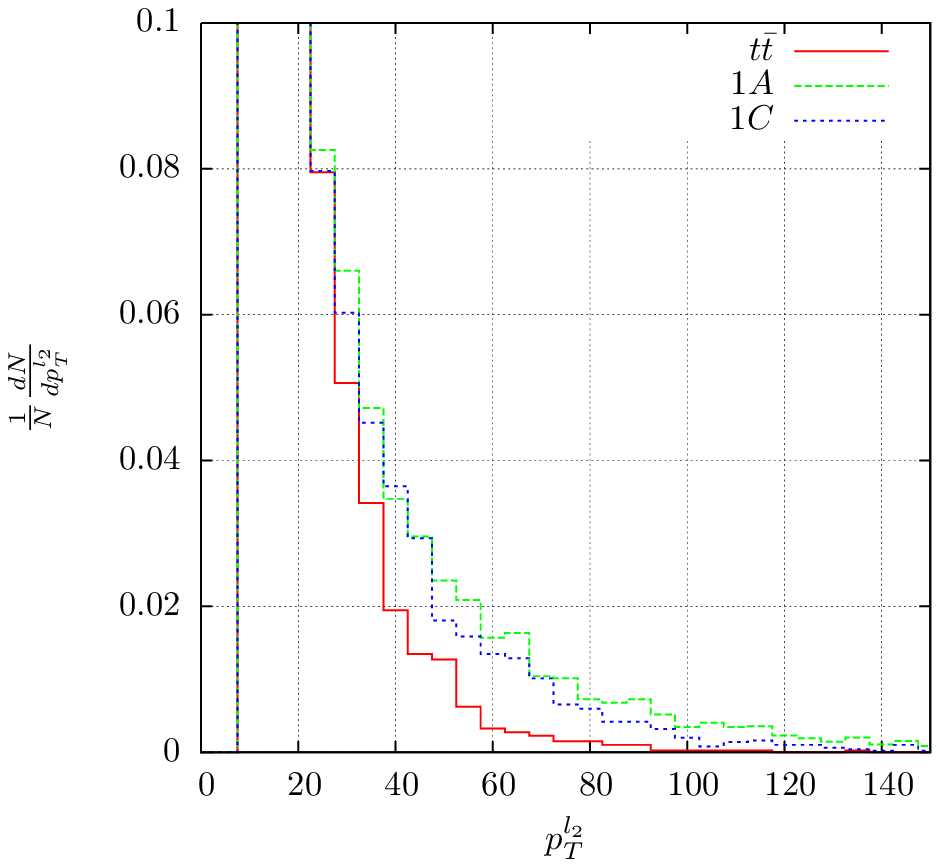}
\caption{Magnitudes of $p_T$ for the two
hardest leptons for points 1A, 1C and standard model $t\bar{t}$ production.}
\end{center}
\end{figure}

\noindent To suppress the $t\bar{t}$ background even further, we
demand that the leptons be of the same sign.  We also look at the
inclusive same-sign dilepton channel (without any b-tags). The signals
and backgrounds for such dilepton events, with and without a tagged
b-jet, are seen in Table~4.  We have calculated the the number of
events, for both signals and backgrounds, corresponding to an
integrated luminosity of $300~\mathrm{fb}^{-1}$.  The advantage of the
di-lepton final states over, say, the $2b$ channel (with or without
one lepton) is quite appreciable.


\begin{table}[tbh]
\begin{center}
\begin{tabular}{|c|c|c|c|c|c|}
  \hline
  \textbf{Point} & \textbf{$1b+2l$} & \textbf{$1b+2l_{(SSD)}$} & \textbf{$2l_{(SSD)}$} & \textbf{$2b+l$} & \textbf{$3b$}\\
  \hline
  \textbf{1A} &15 &6 &25 &4 &2   \\
  \hline
  \textbf{1B} &83 &35 &117 & 27 &24 \\
  \hline
  \textbf{1C} &478 &221 &626 &147 &175  \\
  \hline
  \textbf{2A} &72 &36 &119 &23 &27 \\
  \hline
  \textbf{2B} &486 &166 &568 &181 &161  \\
  \hline
  \textbf{3A} &84 &35 &143 &19 &20 \\
  \hline
  \textbf{3B} &13(5) &109 &592 &243 &712 \\
  \hline
  \textbf{Background} & 10 & 4 & 4 &1514 &5 \\
  \hline
\end{tabular}
\caption{Signals and backgrounds for different channels for an
  integrated luminosity of $300~\mathrm{fb}^{-1}$.}
\end{center}
\end{table}

\subsection{Channels: $2b+l$ and  $3b$}

The first consequence of having only third family squarks accessible
is that all SUSY processes involving the production of strongly
interacting superparticles lead to a multiplicity of b's in the final
state.  As we have mentioned already, most of these have too high
$p_T$ to be reliably tagged.  However, there will still be sufficient
number of events with two or three b-tags. There one has to compromise
on lepton identification, so as to gain in branching ratios. On the
whole, this reflects a tug-of-war between the loss in rate due to
branching ratios and that due to our demand that only b's in a
specific $p_T$-range be identified.  Thus for identifying events with
high squark and gluino masses, where the cross section is already very
low, we recommend looking at only single-lepton events when more than
one b's are tagged.

For two b-tagged events, we find a very large background from
$t\bar{t}$ processes. We suppress this by demanding the presence of a
high-$p_T$, isolated lepton, satisfying $p_T(l_1) \geq
80~\mathrm{GeV}$.  The requirement of leptons has to be given up for
3b events, for otherwise the overall rates will be far too small.

The primary backgrounds for $2b+l$ channel are same as $1b+l$, viz.
$t\bar{t}+jets,Wb\bar{b}+jets,Wt\bar{t}+jets,Zt\bar{t}+jets,Zb\bar{b}+jets,
4t,4b$ and $2t2b$. Again, we have used ALPGEN to compute the background
rates.

Since the $3b$ cannot result from tree-level standard model processes
(excepting those suppressed by weak mixing), the backgrounds are only
due to $4t,2t2b$ and $4b$. However, the $4b$ processes do not have a
source of high $E_T \miss$, so the highest contribution comes from
$2t2b$ production processes.  

The results are presented in Table~4.

\subsection{Inclusion of the Higgs}

In this study, we wish to emphasise situations where the gluino mass
is $\gsim ~ 1~TeV$.  This roughly corresponds the region of the
parameter space with $m_{1/2} \ge 400~GeV$. As can be seen from
Figure~7, decay $\tilde{\chi}_2^0 \rightarrow h \tilde{\chi}_1^0$ has
a branching ratio greater than 90\% over most of the region of
parameter space for $\tan\beta~=~5$.  $\tilde{\chi}_2^0 \rightarrow Z
\tilde{\chi}_1^0$ is suppressed in these regions, and the lightest
neutral Higgs occurs in a significant number of events in this
scenario.  For $\tan\beta~=~40$, this region is much reduced and the
decay into a Higgs is appreciable only in the region $m_\frac{1}{2} >
700~GeV$ where the gluino mass is close to the upper limit of
accessibility.  The dominant decay then is $\tilde{\chi}_2^0
\rightarrow Z \tilde{\chi}_1^0$.  Thus the production of the Higgs can
give information whether $\tan \beta$ is high or low.

\begin{figure}[h]
\begin{center}
\includegraphics[width=70mm]{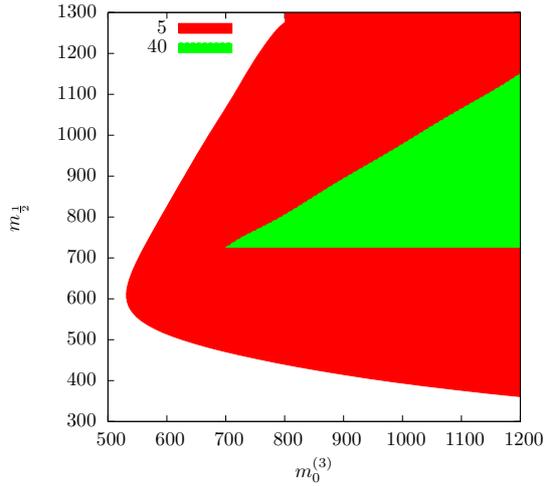}
\caption{Regions in parameter space corresponding to the branching fraction 
$BF(\tilde{\chi}_2^0 \rightarrow h \tilde{\chi}_1^0)>0.9$ for $\tan \beta =5$ and $40$.}
\end{center}
\end{figure}

We are discussing a situation where the lightest neutral Higgs has
already been discovered and it's mass is known. Ideally, one would
like to identify the Higgs by picking a b-jet pair with it's invariant
mass near the mass of the Higgs.  However, in most events, both b's
from the Higgs cannot be identified (as seen from Figure~7).  And
demanding only one b-tagged jet instead of two leads to a combinatorial
background much higher than actual number of signal events.  To be
able to reduce this, we compare the $p_T$ distribution of jets from
Higgs decay and the opening angle between the jets for true Higgs
events and the combinatorial background.  The distributions are shown
in Figure~7.

\begin{figure}[th]
\begin{center}
\includegraphics[width=70mm]{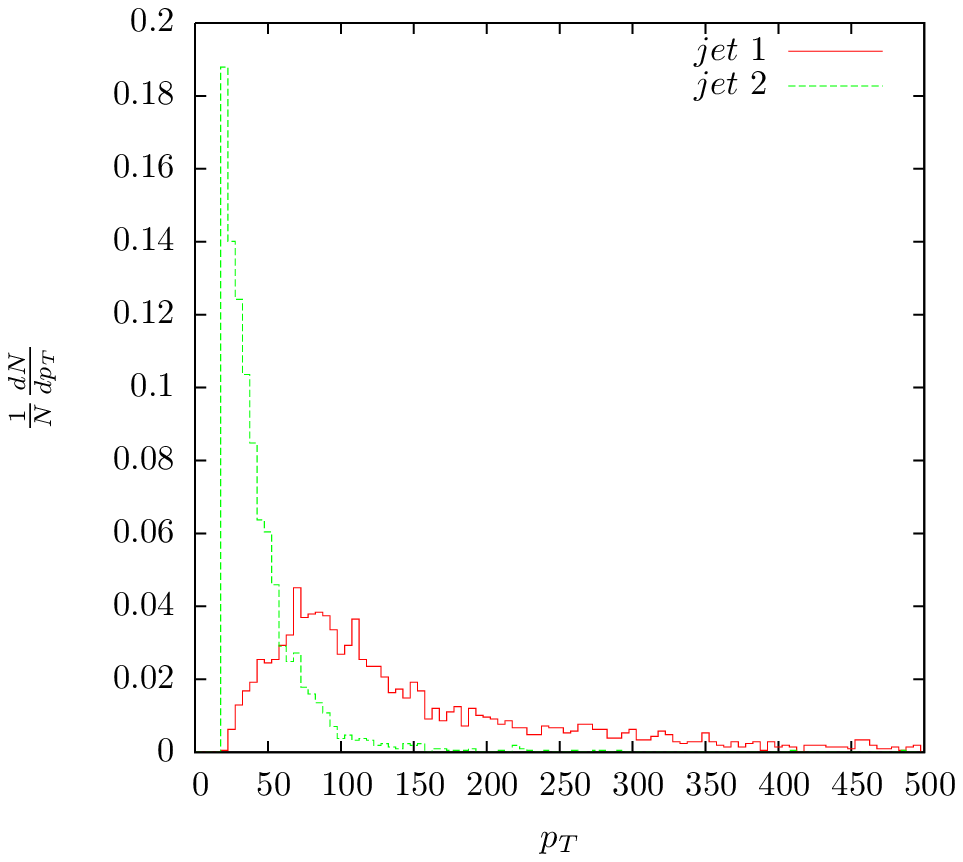}
\includegraphics[width=70mm]{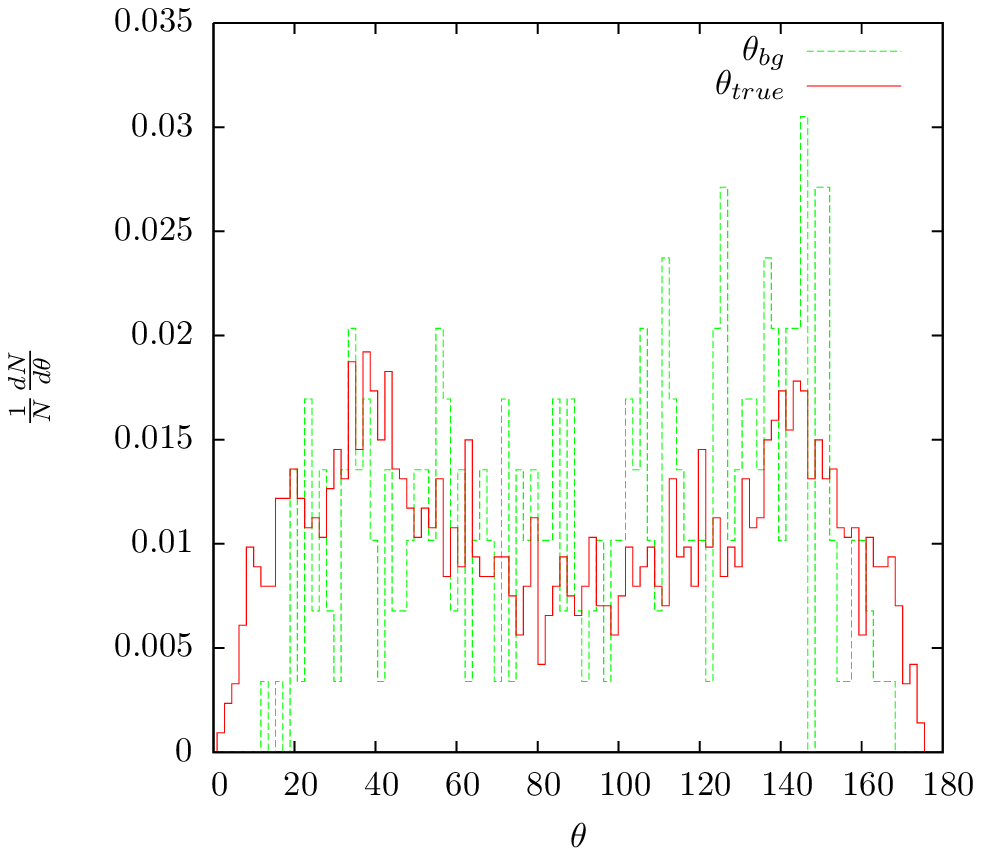}
\caption{The transverse momentum ($p_T$) of jets from Higgs decay and
  the opening angle $\theta$ between the jets for signal events
  (``$\theta_{true}$'') and the combinatorial
  background(``$\theta_{bg}$'').}
\end{center}
\end{figure}

We then claim to  have  identified a Higgs through a jet pair if:
\begin{enumerate}
  \item $|M_{j_1 j_2} - M_{h}| < 15.0~\mathrm{GeV}$ where $M_{j_1
    j_2}$ is the invariant mass of the jet pair.
  \item The second (less energetic) jet has $p_T<80~GeV$.
  \item {\em At least one of the two jets} is b-tagged.
  \item The opening angle between the jets is less than $\pi/2$.
\end{enumerate}

These cuts reduce the combinatorial background to about half that of
the signal.  Identifying the Higgs means at least one b-tag.
Therefore, we study the channels $2l+h$, $2l_{(SSD)}+h$, $1b+l+h$ and
$2b+h$ with exactly the same hard-lepton cuts. The signals and
backgrounds for all Higgs channels are summarised in Table~5.  The
combinatorial background is mentioned in the parenthesis accompanying each
number of signal events. 

We find that for points with gluino mass $\gsim 1.5~TeV$, the event
rates are not significant enough to make a distinction between the
region favouring Higgs production and the region where it is
suppressed.  However, for points 1C, 2B and 3B, we can see a clear
distinction in the number of Higgs events.  In particular, the
$1b+l+h$ and $2b+h$ channels have the added advantage of having a low
combinatorial background.  These channels show a significant excess
even after taking the combinatorial background into consideration.  The
leptonic channels have a large combinatorial background which make them
unreliable for making definite statements about Higgs production with
the identification criteria stated above.  Thus, one can use this
information to infer whether the situation corresponds to low or high
$\tan \beta$.

\begin{table}[tbh]
\begin{center}
\begin{tabular}{|c|c|c|c|c|}
  \hline
  \textbf{Point} & \textbf{$2l+h$} & \textbf{$2l_{(SSD)}+h$} & \textbf{$1b+l+h$} & \textbf{$2b+h$} \\
  \hline
  \textbf{1A} &3(1) &1(0) &1(0) &0 (0)   \\
  \hline
  \textbf{1B} &13 (5) & 5(2) &3(2) &3(0)  \\
  \hline
  \textbf{1C} &110(60) &28(14) &32(9) &37(9)  \\
  \hline
  \textbf{2A} &12(3) &4(1) &5(1) &8(1)  \\
  \hline
  \textbf{2B} &96(55) &40(25) &30(10) &46(5)  \\
  \hline
  \textbf{3A} &13(5) &6(2) &5(2) &3(0)  \\
  \hline
  \textbf{3B} &132(121) &69(69) &0(0) &5(5)  \\
  \hline
  \textbf{Background} & 5 & 3 &7 &3 \\
  \hline
\end{tabular}
\caption{Signals and backgrounds for different channels with Higgs
  identification for an integrated luminosity of
  $300~\mathrm{fb}^{-1}$. The irreducible combinatorial background for
  each channel is given the the parentheses.}
\end{center}
\end{table}

\section{Distinction from scenarios where the first two sfermion families
are also accessible}

While signals have been suggested above for discovering SUSY with only
the third family light, it is also instructive to ask whether such a
scenario can be distinguished from the more frequently discussed case
where all three families are within the reach of the LHC.  We take up
such a discussion in this section, showing that this can be done by
(a) considering the `effective mass' distribution of events, and (b)
taking event ratios for different channels.  For illustration, we
choose the benchmark point 1C from our previous analysis and choose
two points generated in the mSUGRA scenario (i.e. all sfermion masses
now arise from the same $m_0$) as representatives of the case when all
three sfermion families are accessible.

The first point (S1) is generated so as to have low-scale stop and
gluino masses as close to 1C as possible.  As one can see from
Figure~8, this corresponds to a nearly identical $M_{eff}$
distribution.  The second point (S2) was generated to give the similar
number of events at $300~fb^{-1}$ in several channels.  Since in our
previous analysis, we have found $SSD$ to be a clean channel, it has
been used here for illustration. The low-scale masses for
third-generation squarks and gluinos for the two mSUGRA points
corresponding to the point 1C are given, along with the high-scale
values of $(m_0,m_{\frac{1}{2}})$, in Table~6.  The values of
$tan\beta$ and $sign(\mu)$ are chosen to be $10$ and positive
respectively.  The trilinear soft breaking parameter $A$ is set to
zero at high scale.

\begin{table}[htbp]
\begin{center}
\begin{tabular}{|c|c|c|c|c|c|c|c|c|}

  \hline
  \textbf{Point} &  \textbf{$tan \beta$} &\textbf{$m_{1/2}$} & \textbf{$m_0~({m_0}^{(3)})$} & \textbf{$m_{\glue}$} & \textbf{$m_{\tilde{t_1}(\tilde{t_2})}$}  & \textbf{$m_{\tilde{b_1}(\tilde{b_2})}$} & \textbf{$m_{\tilde{u_1}(\tilde{u_2})}$}\\
  \hline
  \textbf{1C} &10 &400 &1200 &1063 &623 (916) &892 (1153) & 5015(5023)\\
  \hline
  \textbf{S1} &10 &400 &100 &998 &697 (895)& 847 (879) &914 (883) \\
  \hline
  \textbf{S2} &10 &570 &1200 &1362 &1163 (1468) & 1453(1616) &1654 (1628)\\
  \hline
\end{tabular}
\caption{Third generation squark and gluino masses in GeV for two
mSUGRA points and point 1C.}
\end{center}
\end{table}

We calculate the event rates for the same channels
($1b+2l$,$1b+2l_{(SSD)}$ ,$2l_{(SSD)}$,$2b+l$,$3b$) as before.  The
  basic cuts as well as any extra cuts applied are same as in
  section~3.  The event rates are given in Table~7.

\begin{table}[tbph]
\begin{center}
\begin{tabular}{|c|c|c|c|c|c|c|}
  \hline
  \textbf{Point} & \textbf{$\sigma_{basic}(fb)$} & \textbf{$1b+2l$} &
  \textbf{$1b+2l_{(SSD)}$} & \textbf{$2l_{(SSD)}$} & \textbf{$2b+l$} &
  \textbf{$3b$}\\
  \hline
  \textbf{1C} &83.87 &478 &221 &626 &147 &175  \\
  \hline
  \textbf{S1} &1160 &1619 &298 &3239 &255 &213   \\
  \hline
  \textbf{S2} &74.63 &446 &195 &622 &117 &123 \\
  \hline
  \textbf{Background}& & 10 & 4 & 4 &1514 &5 \\
  \hline
\end{tabular}
\caption{Number of events at $300~fb^{-1}$ for the mSUGRA points S1
  and S2.  We have repeated the numbers for point 1C and background for
  comparison.}
\end{center}
\end{table}

In R-parity conserving SUSY, only even number of superparticles can be
produced.  Therefore, the peak of the $M_{eff}$ distribution
corresponds roughly to twice the mass of the lightest superparticle
pair-produced through hard scattering.  This gives us an indication of
the mass scale of SUSY particles.  It should be noted that, in
mSUGRA-based models, too, the third family sfermions are usually the
lightest (though the first two are not necessarily decoupled.) Thus
the masses of the gluino and/or the third family squarks will be
indicated by the peak of the $M_{eff}$ distribution.  The $M_{eff}$
distributions for points 1C, S1 and S2 are shown in Figure~8.

\begin{figure}[htbp]
\begin{center}
\includegraphics[width=70mm]{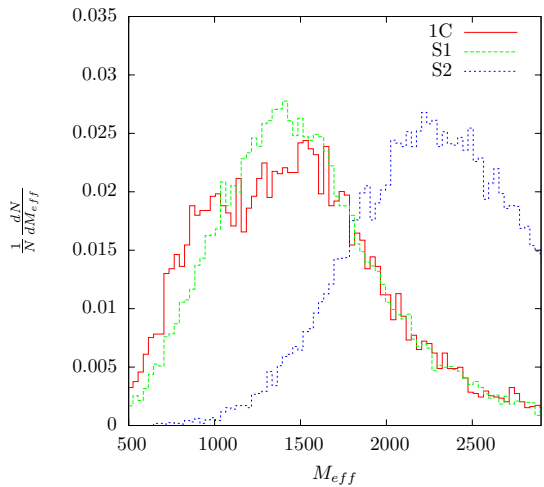}
\caption{Comparison of $M_{eff}$ distributions for points 1C, S1 and S2.}
\end{center}
\end{figure}

Based on the information from the $M_{eff}$ distribution and the event
rates, we can draw the following inferences:
\begin{enumerate}
  \item \begin{enumerate}
    \item The points 1C and S1 have a very similar spectrum for third
      generation squarks and gluino masses.  They are not
      distinguishable by looking at the $M_{eff}$ distribution alone.
    \item The cross section for squark and gluino production for S1 is
      very high since all the squarks are accessible.  Note, in
      particular, that the ratio $3b_{1C}:3b_{S1}=0.82$ is close to
      one whereas the ratio $SSD_{1C}:SSD_{S1}=0.19$ is much smaller.
      The $3b$ final state which comes only from $\tilde{g}\tilde{g}$
      production shows comparable number of events due to similar
      gluino mass.  The masses of $\tilde{t}_2$, $\tilde{b}_1$ and
      $\tilde{b_2}$ are smaller in S1 as compared to 1C.  The b's in
      their cascades are therefore likely to have lower $p_T$ and
      therefore, their identification efficiency will be higher.
    \item Rates for the channels $1b+2l$ and $SSD$ are highly enhanced
      for the points S1.  Since $\tilde{q}\tilde{q}$,($q=u,d,s,c$) is
      allowed, their cascades into charginos yield larger number of
      dileptons.  This also explains why on demanding one b-tag
      ($1b+2l(SSD)$ channel), the increase in the number of events is
      not so dramatic.
  \end{enumerate}

  \item \begin{enumerate}
    \item The $M_{eff}$ distribution for the two points is very
      different and easily distinguishable.
    \item As intended, the number of events in the $SSD$ channel are
      nearly same $3b_{1C}:3b_{S2}=1.01$ for points 1C and S2.  The
      $3b$ channel however, shows more events in the case of 1C
      ($SSD_{1C}:SSD_{S2}=1.42$.)  This is to be expected since the
      mass of the gluino is higher for S2 and therefore, the cross
      section of $\tilde{g}\tilde{g}$ is lower. Also, the masses of
      $\tilde{t}_{1,2}$ and $\tilde{b}_{1,2}$ are higher resulting in
      higher $p_T$ of b's in the final state and hence lower
      identification efficiency.
  \end{enumerate}
\end{enumerate}

Thus we find the the total cross sections for sparticle production are
much lower for the case where only third family sfermions are
accessible, making detection more challenging than the case where all
three generations have masses $\sim 1~TeV$.  However, the points in
parameter space of mSUGRA which mimic the scenario are characterised
either by a very different effective mass distribution or very
different rates in the leptonic channels.  We can conclude that
this scenario can be distinguished from a universal scenario with all
three generations are accessible.

\section{Summary and conclusions}
We have investigated the signals of SUSY at the LHC, when only the
third squark family is kinematically accessible. We have emphasised
the difficulties in identifying highly energetic tops and bottoms and
suggested various combinations of b-and leptonic final states,
including those with like-sign dileptons as viable alternatives to
reconstruction of the top.  Only those b's whose $p_T$ lies in the
range $50-100$~GeV have been included so that the tagging efficiency
is optimal.  We have also used a large missing-$E_T$ cut of
$300~\mathrm{GeV}$, and taken particular care in calculating the
missing energy, including soft contributions to the visible energy.
The above event selection criteria, together with the variable
effective mass, become particularly useful is eliminating
backgrounds. Also, the fact that such scenarios have large production
of the lightest neutral Higgs on-shell (in the decay $\chi^0_2
\rightarrow \chi^0_1 h$) when $\tan \beta$ is low gives us an
additional handle on identifying the order of $\tan \beta$.

There are earlier studies in similar directions, to which we have
already referred. In addition, while this paper was almost complete,
we came to know about another work~\cite{Acharya:2009gb} where studies
in similar lines have been carried out. While we agree with their main
points, we have gone beyond  the parameter region used by them 
(with $m_{\tilde{g}}=650~GeV$), and have explored the regions
where gluinos are close to the LHC search limit, thus addressing
relatively `difficult' regions in the parameter space where
event rates are low.  

We end by re-iterating that the present work has improved upon each of
the earlier ones in the following respects. (a) The difficulties in
identifying high-energy tops and bottoms have been explicitly addressed
(b) The squark and gluino production cross section for such scenarios
is one to two orders of magnitude lower than the universal case with
all three families accessible.  Even with limited top and
b-identification, our multichannel analysis, {\em strengthened by the
  use of leptons having specified kinematic properties in the final
  state}, can take the discovery reach at the LHC for such scenarios
to close to $2~\mathrm{TeV}$ in the gluino mass, for an integrated
luminosity of $300~\mathrm{fb}^{-1}$.  (c) The suggestion of using the
associated Higgs production is of added advantage, as it emphasises
the nature of the spectrum through the viability of the decay
$\chi^0_2 \rightarrow \chi^0_1 h$ depending on the value of $\tan
\beta$.  (d) The prospect of {\em distinguishing the scenario under
  investigation from one with all three sfermion families accessible
  is emphasised through a combination of kinematic studies and ratios
  of event rates in various channels}. Thus it is hoped that not only
can one discover a SUSY scenario where only the third family is
accessible, but can also set the scenario apart from the ones with all
scalar families accessible, when sufficient luminosity accumulates at
the LHC.

{\bf Acknowledgement:} We thank Bruce Mellado for a number of useful
suggestions.  In addition, we acknowledge helpful discussions with
Priyotosh Bandyopadhyay, Subhaditya Bhattacharya, Sanjoy Biswas,
Aseshkrishna Datta and V. Ravindran.  This work was partially
supported by funding available from the Department of Atomic Energy,
Government of India for the Regional Centre for Accelerator-based
Particle Physics, Harish-Chandra Research Institute. Computational
work for this study was partially carried out at the cluster computing
facility of Harish-Chandra Research Institute ({\tt
http:/$\!$/cluster.mri.ernet.in}).

\end{document}